\documentstyle[bezier,epsfig,12pt,preprint,tighten,aps]{revtex}
\begin{document}

\draft

\title{\rightline{{\tt (January 1998)}}
\rightline{{\tt TMUP-HEL-9801}}
\rightline{{\tt UM-P-98/04}}
\rightline{{\tt RCHEP-98/01}}
\ \\
Comparing and contrasting the $\nu_{\mu} \to \nu_{\tau}$
and $\nu_{\mu} \to \nu_s$ solutions to\\
the atmospheric neutrino problem with SuperKamiokande data}
\author{R. Foot and R. R. Volkas}
\address{School of Physics\\
Research Centre for High Energy Physics\\
The University of Melbourne\\
Parkville 3052 Australia\\
(foot@physics.unimelb.edu.au, r.volkas@physics.unimelb.edu.au)}
\author{O. Yasuda}
\address{Department of Physics\\
Tokyo Metropolitan University\\
1-1 Minami-Osawa Hachioji, Tokyo 192-03, Japan\\
(yasuda@phys.metro-u.ac.jp)}
\maketitle

\begin{abstract} The $\nu_{\mu} \to \nu_{\tau}$ and $\nu_{\mu} \to \nu_s$
solutions to the atmospheric neutrino problem are compared with
SuperKamiokande data. The differences between these solutions due to matter
effects in the Earth are calculated for the ratio of $\mu$-like to $e$-like
events and for up-down flux asymmetries. These quantities are chosen
because they are relatively insensitive to theoretical uncertainties in the
overall neutrino flux normalisation and detection cross-sections and
efficiencies. A $\chi^2$ analysis using these quantities is performed
yielding $3\sigma$ ranges which are approximately given by $(0.725 - 1.0,\
4 \times 10^{-4} - 2 \times
10^{-2}\ eV^2)$ and $(0.74 - 1.0,\ 1 \times 10^{-3} - 2 \times 10^{-2}\
eV^2)$ for $(\sin^2 2\theta,\Delta m^2)$ for the $\nu_{\mu} \to \nu_{\tau}$
and $\nu_{\mu} \to \nu_s$ solutions, respectively. Values of $\Delta m^2$
smaller than about $2 \times 10^{-3}$ eV$^2$ are disfavoured for the
$\nu_{\mu} \to \nu_s$ solution, suggesting that future long baseline
experiments should see a positive signal if this scenario is the correct
one. \end{abstract}

\newpage

Atmospheric neutrino data provides important evidence for the existence of
neutrino oscillations \cite{atmos,sk}. Atmospheric neutrinos are produced
primarily from
the decays of mesons and muons which result from interactions between the
primary cosmic ray flux and air molecules in the Earth's upper atmosphere. 
The neutrino flux consists mostly of $\nu_e$, $\overline{\nu}_e$,
$\nu_{\mu}$ and $\overline{\nu}_{\mu}$, with the muon flavour flux expected
to be roughly twice as large as the electron flavour flux. When these
neutrinos interact with matter via the charged current they produce the
corresponding charged leptons or antileptons.  Atmospheric neutrino
experiments measure the corresponding event rates.  Historically, the
atmospheric neutrino anomaly has been defined to be the discrepancy between
measured values of the ratio of $\mu$-like to $e$-like events ($\sim 1.2$) 
to the predicted ratio of about 2. The SuperKamiokande experiment has
recently produced relatively high statistics data that considerably
strengthens the case for an atmospheric neutrino anomaly \cite{sk}. In
particular,
the new data provide strong evidence for an anomalous zenith-angle
dependence for multi-GeV $\mu$-like events. Furthermore, the pattern of
this dependence is consistent with a neutrino oscillation
explanation.

Generically, the atmospheric neutrino anomaly points for its solution
towards large angle $\nu_{\mu} \to \nu_e$ \cite{mue}, $\nu_{\mu} \to
\nu_{\tau}$ \cite{mutau} or $\nu_{\mu} \to \nu_s$ \cite{mus} oscillations,
where $\nu_s$ denotes a sterile neutrino \cite{3fl}. However, $\nu_{\mu}
\to \nu_e$ is now disfavoured (though not completely ruled out)  by results
from the CHOOZ reactor-based $\overline{\nu}_e$ disappearance experiment
\cite{chooz}. We therefore focus on the $\nu_{\mu} \to \nu_{\tau}$ and
$\nu_{\mu} \to \nu_s$ possibilities in this paper. A major goal of future
atmospheric neutrino research should be to discriminate between these rival
solutions. To this end, the purpose of this paper is to compare and
contrast these two most favoured oscillation modes with the latest
SuperKamiokande data. 

Though in many respects similar, the $\nu_{\mu} \to \nu_{\tau}$ and
$\nu_{\mu} \to \nu_s$ cases are distinguishable, because $\nu_{\tau}$
interacts via the neutral current with ordinary matter whereas $\nu_s$, by
definition, does not. Neutral current interactions with the Earth affect
the evolution of the $\nu_{\mu} + \nu_s$ system because of the matter
effect \cite{msw}. The evolution of the $\nu_{\mu} + \nu_{\tau}$ system is,
by contrast, identical to what it would be in vacuum. A major goal of this
paper is study the magnitude of the matter effect. We will show that it is
quite important for multi-GeV events if $\Delta m^2 < 10^{-2}$ eV$^2$ and
for sub-GeV events if $\Delta m^2 < 10^{-3}$ eV$^2$. We will also perform a
$\chi^2$ analysis of the two cases with respect to the SuperKamiokande
data. It is interesting to note that other ways of discriminating between
$\nu_{\tau}$ and $\nu_s$ have been suggested in the literature. Reference
\cite{vissani} discusses how the sensitivity of SuperKamiokande to
neutral current interactions can be used, while Ref.\cite{liu}
discusses the importance of the matter effect for high energy neutrinos
that
produce upward-going muons. Intriguingly, the MACRO experiment
\cite{macro} sees dips in
the upward-going muon flux at zenith angles that are qualitatively
consistent with expectations from Ref.\cite{liu}. The two cases may also
be distinguished in the future using long baseline experiments,
either through the sensitivity of the detector to neutral currents, or,
for higher energy experiments, by searching for $\nu_{\tau}$ appearance.

In the water-Cerenkov SuperKamiokande experiment,
neutrinos are detected via the charged leptons $\alpha$~
($\alpha$ = $e$ or $\mu$) produced from 
neutrino scattering off nucleons in
the water molecules: $\nu_\alpha N \rightarrow \alpha X$, 
where the identity of $X$ will be discussed below. 
The total number $N(\alpha)$ of charged leptons of
type ${\alpha}$ produced in either of the two scenarios considered is given
by
\begin{eqnarray}
\displaystyle
N(\alpha)
&=& n_T \
\int_0^\infty dE
\int^{q_{\rm max}}_{q_{\rm min}} dq
\int_{-1}^{+1} d\cos \psi
\int_{-1}^{+1} d\cos \xi\ 
{1 \over 2\pi}
\int_{0}^{2\pi} d\phi 
\nonumber\\
&\times&
{d^2F_\alpha (E,\xi) \over dE~d\cos\xi}
\cdot{ d^2\sigma_\alpha (E,q,\cos\psi) \over dq~d\cos\psi }
\cdot
{\ }P(\nu_\alpha\rightarrow\nu_\alpha; E, \xi).
\label{rate}
\end{eqnarray}
Here $d^2F_\alpha /dEd\cos\xi$ is the differential flux of atmospheric
neutrinos of type $\nu_\alpha$ of energy $E$ 
at zenith angle $\xi$.
The term $n_T$ is the effective number of target nucleons.  
The function $d^2\sigma_\alpha/dqd\cos\psi$ is the differential cross section
for $\nu_\alpha N \rightarrow \alpha X$ scattering,
where $q$ is the energy of the charged lepton and
$\psi$ is the scattering
angle relative to the velocity vector of the incident $\nu_{\alpha}$ (the
azimuthal angle having been integrated over). The function
$P(\nu_\alpha\rightarrow\nu_\alpha; E, \xi)$ is the survival
probability for a $\nu_\alpha$ with energy $E$ after
travelling a distance
$\displaystyle L=\sqrt{(R+h)^2-R^2\sin^2\xi}-R\cos\xi$,
where $R$ is the radius of the Earth and $h\sim$ 15 km is the mean altitude
at which atmospheric neutrinos are produced.
Finally note that $\phi$ is the azimuthal angle relative to the incident
neutrino direction (see Fig.1 for an illustration of all the relevant
angles).
The integration over $\phi$ is only non-trivial when calculating 
zenith angle binned data. In order to calculate the 
number of $\alpha$-like events for certain
energy ranges and within certain zenith angle bins, 
the integration ranges in Eq.(\ref{rate}) must be
truncated accordingly, with the direction of the charged lepton then
obtained from
\begin{equation}
\cos \Theta = \cos \xi \cos \psi + \sin \xi \cos \phi \sin \psi,
\end{equation}
where $\Theta$ is the zenith angle of the charged lepton.

Since not all charged current events are used in the SuperKamiokande
analysis, Eq.\ref{rate} must be modified to incorporate the so-called
detection efficiency function.
This function is defined by
\begin{equation}
{\it detection \ efficiency \equiv 
{Number \ of \ 1 \ ring \ charged \ current \ events \over Total \
 number \ of\  charged \ current \ events}\ .}
\end{equation}
For our sub-GeV analysis we use the approximation of only including
quasi-elastic scattering. In this approximation the detection
efficiency function is set to 1 because quasi-elastic scattering 
always lead to 1 ring events. In the multi-GeV 
analysis, we use a detection efficiency function obtained from the
SuperKamiokande collaboration.

The survival probability $P(\nu_\alpha\rightarrow\nu_\alpha; E, \xi)$ is
obtained by solving the
Schr\"odinger equation for neutrino evolution including matter effects. 
It is given by
\begin{equation}
i {d \over dx} \left[ \begin{array}{c} \nu_{\mu}(x) \\ \nu_{\tau,s}(x)
\end{array} \right] = 
\left[ \begin{array}{cc}
{\Delta m^2 \over 2E}\sin^2 \theta\  & {\Delta m^2 \over
2E}\sin\theta\cos\theta \\
{\Delta m^2 \over 2E}\sin\theta\cos\theta\  &
{\Delta m^2 \over 2E}\cos^2 \theta + A_{\tau,s}(x) \end{array} \right]
\left[ \begin{array}{c} \nu_{\mu}(x) \\ \nu_{\tau,s}(x)
\end{array} \right],
\end{equation}
where $x$ is the distance travelled, $\Delta m^2$ the difference in squared
masses, $\theta$ the vacuum mixing angle and $\nu_{\mu,\tau,s}(x)$ the
wave-functions of the neutrinos. The quantities $A_{\tau,s}(x)$
are the effective potential differences generated through the matter effect:
\begin{equation}
A_{\tau}(x) = 0
\end{equation}
and, for electrically neutral terrestrial matter\cite{bla}
\begin{equation}
A_s(x) = {\sqrt{2} \over 2} G_F N_n(x) = {\sqrt{2} \over 2} G_F (Y_n/m_n)
\rho(x),
\end{equation}
where $G_F$ is the Fermi constant, $N_n(x)$ is the number density of
neutrons along the path of the neutrino, $Y_n \simeq 0.52$ is the average
number of
neutrons per nucleon, $m_n$ is the nucleon mass and $\rho(x)$ is the
mass density. Our numerical
calculations use the density profile of the Earth given in
Ref.\cite{earth}.
The $\nu_{\mu}$ survival probability is given by
$\nu_{\mu}(L)^{*}\nu_{\mu}(L)$. For antineutrinos the sign of $A_s$ is
reversed.

The differential flux of atmospheric neutrinos
$d^2F_\alpha /dEd\cos\xi$ without geomagnetic effects
is given in \cite{hkkm}, but we have used the differential
flux which includes geomagnetic effects \cite{hkm}. (For other atmospheric
neutrino flux calculations, see Ref.\cite{flux}.)  

SuperKamiokande separates its data into a sub-GeV sample and a multi-GeV
sample. For sub-GeV events, charged leptons are dominantly produced via
quasi-elastic scattering: $\nu_{\alpha} N \to \alpha N'$, where $N$ and
$N'$ are nucleons. The state
$X$ [see Eq.(\ref{rate})] is therefore identified with $N'$ for these
events. We use the cross-section given in Ref.\cite{gaisser}. The struck
nucleon $N$
is either a proton in hydrogen, or a bound
nucleon in oxygen. In the case of hydrogen, the 2-body nature of
quasi-elastic scattering leads to a relation between $E$, $q$ and $\psi$
from relativistic energy-momentum conservation. One of the integrations in
Eq.(\ref{rate}) is therefore redundant for scattering off hydrogen. For
scattering off a nucleon within
an oxygen nucleus, Fermi motion and Pauli blocking effects are incorporated
via the
prescription in Ref.\cite{gaisser}. In this case, there is no relation
between $E$,
$q$ and $\psi$ because of the nuclear effects.
For, multi-GeV events we use the inclusive cross-section for $\nu_{\alpha} N \to
\alpha X$ given in Ref.\cite{barger}. 
Although this cross section is not completely satisfactory for
calculating absolute event rates because
it does not incorporate low $Q^2$ effects such as $\Delta$ resonance
production, it is a good enough approximation for calculating ratios of
event rates such as
up-down asymmetries \cite{updown} and $N(\mu)/N(e)$ because 
these quantities are relatively insensitive to details.
Indeed the main advantage in using event rate ratios 
is that they are relatively insensitive to 
uncertainties in the cross-sections and the neutrino fluxes.
In this way $20-30\%$ uncertainties in
overall flux normalisations and cross-sections are avoided in favour of
quantities with systematic uncertainties of only a few percent. 

We will now define the event rate ratios used in the analysis.
We first define the traditional quantity R, where
\begin{equation} 
R \equiv \frac{(N_{\mu}/N_e)|_{osc}}{(N_{\mu}/N_e)|_{no-osc}}.  
\end{equation}
The quantities $N_{e,\mu}$ are the numbers of $e$-like and 
$\mu$-like events, as per Eq.(\ref{rate}). 
The numerator denotes numbers obtained from Eq.(\ref{rate}), while the
denominator the numbers expected with oscillations switched off. A class of 
up-down flux asymmetries for $\alpha$-like events is defined by
\begin{equation}
Y^{\eta}_{\alpha} \equiv {(N_{\alpha}^{-\eta}/N_{\alpha}^{+\eta})|_{osc}
\over (N_{\alpha}^{-\eta}/N_{\alpha}^{+\eta})|_{no-osc}}.
\end{equation}
Here $N_{\alpha}^{-\eta}$ denotes the number of $\alpha$-like events 
produced in
the detector with zenith angle $\cos \Theta < -\eta$, while
$N_{\alpha}^{+\eta}$ denotes the analogous quantity for $\cos \Theta >
\eta$, where $\eta$ is defined to be positive. SuperKamiokande divides the
$(-1,+1)$ interval in $\cos\Theta$ into five equal bins. The central bin
straddles both the upper and lower hemispheres, and is thus not useful for
up-down asymmetry analyses. We therefore choose $\eta = 0.2$ in order to
utilise all the data in the other four bins. Since $\nu_e$'s do not
oscillate in the two scenarios $\nu_{\mu} \to \nu_{\tau,s}$ which
we consider, up-down asymmetries for $e$-like events, $Y^{0.2}_e$ are predicted 
to equal 1. Note that systematic uncertainties for up-down asymmetries are
smaller than for $R$, because the latter depends on the relative flux of
$\nu_{\mu}$ to $\nu_e$.

In the context of the $\nu_{\mu} \to \nu_{\tau,s}$ scenarios considered
here, $R$ measures the disappearance of $\nu_{\mu}$'s and
$\overline{\nu}_{\mu}$'s relative to $\nu_e$'s and $\overline{\nu}_e$'s and
to no-oscillation expectations. The up-down asymmetries probe the zenith
angle dependences of the neutrino fluxes relative to no-oscillation
expectations. Both classes of quantities provide pertinent information
about the pattern of the putative $\nu_{\mu}$ oscillations
while being insensitive to the systematic uncertainties discussed above.

Additional important information about the oscillation pattern is supplied
by the energy dependences of $R$ and the $Y$'s. For this reason we
calculate separate $R$'s and $Y$'s for the SuperKamiokande sub-GeV and
multi-GeV samples. The sub-GeV sample is defined by the momentum cuts $0.1
< p_e/GeV < 1.33$ and $0.2 < p_{\mu}/GeV < 1.5$ for $e$-like and $\mu$-like
events, respectively. We also consider an alternative low-energy cut which
has a lower limit of $0.5$ GeV. The alternative cut enhances the effect of
oscillations because the correlation of the produced charged leptons with the
incident neutrinos is stronger for higher energies.

Our results for the $R$'s and $Y$'s are displayed in Figs.2-6
\cite{fn} together with
the preliminary SuperKamiokande results \cite{sk},
\begin{eqnarray}
R ({\it sub-GeV}) & = & 0.61 \pm 0.03 \pm 0.05, \nonumber\\
R ({\it multi-GeV}) & = & 0.67 \pm 0.06 \pm 0.08, \nonumber\\
Y^{0.2}_{\mu} ({\it sub-GeV}) & = & 0.78 \pm 0.06, \nonumber\\
Y^{0.2}_{\mu} ({\it multi-GeV}) & = & 0.49 \pm 0.06. 
\end{eqnarray}
The preliminary experimental results we use correspond to 414 live days of
running.
Note that experimental results for the alternative sub-GeV sample 
with $p_{e,\mu} > 0.5$ GeV are not at present available.
For completeness
we mention that the preliminary SuperKamiokande results for the
$e$-like up-down asymmetries are
\begin{eqnarray}
Y^{0.2}_{e} ({\it sub-GeV}) & = & 1.13 \pm 0.08, \nonumber\\
Y^{0.2}_{e} ({\it multi-GeV}) & = & 0.83 \pm 0.13.
\end{eqnarray}
Finally note that only statistical errors are given for the up-down
asymmetries since they should be much larger than possible systematic
errors.

These results have several significant features: (i) For the sub-GeV cases,
the matter effects become noticeable at about $\Delta m^2 = 10^{-3}$ eV$^2$
and are really very significant at $\Delta m^2 = 10^{-4}$ eV$^2$. (ii) The
matter effects cut in at the higher value of about $\Delta m^2 = 10^{-2}$
eV$^2$ for the multi-GeV cases. (iii) The up-down muon asymmetries plateau
between about $10^{-3}$ and $10^{-2}$ eV$^2$, with the sub-GeV plateau
occuring for slightly lower values of $\Delta m^2$ compared to the
multi-GeV case. For this range of $\Delta m^2$, downward travelling
neutrinos do not have time to oscillate, whereas upward travelling
neutrinos experience averaged oscillations. The plateau phenomenon provides
a characteristic prediction for up-down asymmetries that is reasonably
insensitive to $\Delta m^2$, while remaining sensitive to the mixing angle.
Note also that the multi-GeV $R$ flattens out in this $\Delta m^2$ range,
for exactly the same reason, with the corresponding plateau for the sub-GeV
$R$ at lower $\Delta m^2$'s (except that the matter effect destroys the
plateau for the $\nu_{\mu} - \nu_s$ scenario). (iv)  Both the sub- and
multi-GeV $Y$'s fit the SuperKamiokande data well for $\Delta m^2$ in the
range $10^{-3} - 10^{-2}$ eV$^2$. The $R$ values also fit the data well in
this range, but with a preference for higher $\Delta m^2$ values. (v) While
a fairly large range of mixing angle values is consistent with both of the
$R$ measurements and the sub-GeV $Y$ datum, the multi-GeV $Y$ result tends
to favour maximal mixing. The multi-GeV $R$ measurement also tends to
favour a large mixing angle. (vi) A comparison of Figs.4 and 5 shows that
the alternative low-energy cut increases both the up-down asymmetry effect
and the matter effect. 

We now perform a $\chi^2$ fit to these data in order to quantify the
significance of the various competing influences discussed above. We define
the $\chi^2$ function as
\begin{equation}
\chi^2 = \sum_E \left[\left({R^{SK} - R^{th} \over \delta R^{SK}}\right)^2
+ \left({Y^{SK}_{\mu} - Y^{th}_{\mu} \over \delta Y^{SK}_{\mu}}\right)^2
+ \left({Y^{SK}_{e} - Y^{th}_{e} \over \delta Y^{SK}_{e}}\right)^2
\right],
\end{equation}
where the sum is over the sub-GeV and multi-GeV cases, the measured
SuperKamiokande values and errors are denoted by the superscript ``SK''
and the theoretical predictions for the quantities are labelled by ``th''.
The $\eta = 0.2$ choice is understood for the up-down asymmetries. We
include both the $e$-like and the $\mu$-like up-down asymmetries in the
fit. There are 6 pieces of data in $\chi^2$ and 2 adjustable parameters,
$\Delta m^2$ and $\sin^2 2\theta$, leaving 4 degrees of freedom.

The statistical procedure we employ is approximate in the sense that ratios
of Gaussian-distributed quantities are only approximately Gaussian
themselves \cite{fogli}. The validity of using ratios increases as the
fractional errors decrease. Since SuperKamiokande is a high statistics
experiment, our procedure is accurate within a sufficiently small region
around the best fit point provided this point gives a good fit 
(we estimate that it is approximately valid within the $3 \sigma$
region around the best fit).  

An alternative $\chi^2$ analysis can be performed by
using absolute
event rates rather than ratios. However in that case the numerical validity
of the results is limited by the correctness of the cross-sections used.
These analyses typically incorporate a $20-30\%$ uncertainty in the event
rates due to uncertain fluxes by introducing theoretical errors in addition
to measurement errors. Unfortunately, existing analyses do not address the
issue of uncertain cross-sections. Our analysis avoids this problem, and is
therefore complementary to the absolute event rate type of analysis. Our
work also extends other recent fits \cite{sk,fit} 
by considering the $\nu_{\mu} \to \nu_s$ case. 

The results of the $\chi^2$ fits are displayed in Figs.7-12. 
Figure 7 shows
the allowed region of $(\sin^2 2\theta,\ \Delta m^2)$ 
at various confidence levels 
for the $\nu_{\mu} \to \nu_{\tau}$ scenario. 
Maximal mixing provides the best fit, and $\Delta m^2$ values in the
$10^{-3}$ to $10^{-2}$ eV$^2$ range are favoured. 
Note that the confidence levels are defined in the usual way by
\begin{equation}
\chi^2 = \chi^2_{min} + \Delta \chi^2
\end{equation}
where $\Delta \chi^2 = 2.3, 4.6, 6.2, 11.8$ for the 
$1\sigma$, $90\%$ C.L., $2\sigma$ and 
$3\sigma$ allowed region respectively.
Our $\chi^2_{min}$ for $\nu_{\mu} \to \nu_{\tau}$ 
oscillations is $\chi^2_{min} = 4.5$
for $4$ degrees of freedom. This is quite a good fit to the data (allowed
at the $35\%$ level).

In Figure 8 we show the allowed region considering just
the asymmetries instead of using both the asymmetries and 
the $R$ ratios. This is of interest because systematic uncertainties
for $Y$'s are smaller than those for $R$'s.
Note that in this case there are 4 data points
and 2 free parameters which gives 2 degrees
of freedom.

According to Fig.9,
$\chi^2$ does not experience a deep minimum at the best fit point
with respect to $\Delta m^2$. This reflects the plateau phenomenon
discussed earlier, and shows that this type of atmospheric neutrino
analysis
will not be able to pinpoint $\Delta m^2$ very precisely. Note that the
minimum becomes shallower when the $R$'s are excluded from the fit. This is
because the plateau in $R$ is not as pronounced as that in $Y$. 

Figures 10-12 show the corresponding results for the $\nu_{\mu} \to \nu_s$
scenario. Smaller values of $\Delta m^2$ are disfavoured in this case
because the matter effect moves both $R$ and $Y$ away from the measured
values.  However, an order of magnitude spread in $\Delta m^2$ is
nonetheless permitted at the $3\sigma$ level. It is interesting to note
that present data tend to predict a positive signal for future long
baseline experiments for the $\nu_{\mu} \to \nu_s$ scenario, whereas the
$\nu_{\mu} \to \nu_{\tau}$ scenario permits $\Delta m^2$ values that are
too small to be probed in this manner.
The value of $\chi^2_{min}$ for the $\nu_{\mu} \to \nu_{s}$ 
scenario is $\chi^2_{min} = 5.1$
for $4$ degrees of freedom. This is similar to 
$\nu_{\mu} - \nu_{\tau}$ case and
also represents quite a good fit (which is 
allowed at $28\%$).

In conclusion, we have demonstrated that matter effects in the Earth have a
significant role to play in comparing and contrasting the $\nu_{\mu} \to
\nu_{\tau}$ and $\nu_{\mu} \to \nu_s$ solutions to the atmospheric neutrino
anomaly with SuperKamiokande data. The matter effects increase both the
ratio of $\mu$-like to $e$-like events, and $\mu$-type up-down asymmetries,
for the $\nu_{\mu} \to \nu_s$ case relative to the $\nu_{\mu} \to
\nu_{\tau}$ case for sufficiently small values of $\Delta m^2$ ($<10^{-2}$
and $<10^{-3}$ eV$^2$ for the multi-GeV and sub-GeV samples, respectively).
Smaller $\Delta m^2$ values ($\stackrel{<}{\sim} 2 \times 10^{-3}$
eV$^2$) are disfavoured for the $\nu_{\mu} \to \nu_s$ scenario, with
interesting implications for future long baseline experiments.

\acknowledgments{O.Y. would like to acknowledge the hospitality of the
School
of Physics at The University of Melbourne where this work was done. O.Y.
was supported in part by a Grant-in-Aid for Scientific Research of the
Ministry of Education, Science and Culture, \#09045036. O.Y. would like to
thank T. Kajita for useful communications and the participants of the
Neutrino Symposium at Hachimantai, Japan, on Nov.28-30 1997 for
discussions. R.F. and R.R.V. are supported by the Australian Research
Council.}

\newpage

\centerline{\large \bf Figure Captions}

\noindent
Figure 1.\ \ The parameterisation of angles in the interaction
$\nu_{\alpha} N \to \alpha X$.

\vspace{5mm}

\noindent
Figure 2.\ \ The sub-GeV $R$ as a function of $\Delta m^2$ for various
values of $\sin^2 2\theta$. The usual SuperKamiokande momentum cuts have 
been employed. The solid (dashed) lines
pertain to the $\nu_{\mu} \to \nu_{\tau}$ ($\nu_{\mu} \to \nu_s$) scenario.
Going from the top to the bottom curves, $\sin^2 2\theta$ takes the values
0.7, 0.8, 0.9 and 1.
Note the significance of the matter effect for $\Delta m^2 < 10^{-3}$
eV$^2$. The dashed-dotted lines denote the preliminary SuperKamiokande
result within a $\pm 1\sigma$ band after 414 live days of running.

\vspace{5mm}

\noindent 
Figure 3.\ \ The multi-GeV $R$ as a function of $\Delta m^2$. Notation as
for Fig.2. Note the significance of the matter effect for $\Delta m^2 <
10^{-2}$ eV$^2$.

\vspace{5mm}

\noindent
Figure 4.\ \ The up-down $\mu$-type asymmetry $Y^{0.2}_{\mu}$ as a function
of $\Delta m^2$ for the sub-GeV sample with the usual SuperKamiokande
momentum cuts. Notation as for Fig.2, except that in the $Y>1$ region the
order of the $\sin^2 2\theta$ values is reversed. Note the significance of
the matter effect for $\Delta m^2 < 10^{-3}$ eV$^2$.

\vspace{5mm}

\noindent
Figure 5.\ \ The up-down $\mu$-type asymmetry $Y^{0.2}_{\mu}$ as a function
of $\Delta m^2$ for the sub-GeV sample with the alternative lower limit
$p_{\mu} > 0.5$ GeV. Notation as for Fig.2, except that in the $Y>1$ region the
order of the $\sin^2 2\theta$ values is reversed. Note the significance of the
matter effect for $\Delta m^2 < 10^{-3}$ eV$^2$.

\vspace{5mm}

\noindent 
Figure 6.\ \ The up-down $\mu$-type asymmetry $Y^{0.2}_{\mu}$ as a function
of $\Delta m^2$ for the multi-GeV sample. Both fully-contained and
partially-contained events are included. Notation as for Fig.2. Note the
significance of the matter effect for $\Delta m^2 < 10^{-2}$ eV$^2$.

\vspace{5mm}

\noindent
Figure 7.\ \ The allowed region in the $(\sin^2 2\theta,\ \Delta m^2)$
plane for the $\nu_{\mu} \to \nu_{\tau}$ scenario.

\vspace{5mm}

\noindent
Figure 8.\ \ As for Fig.7 but with $R$ data excluded from the fit.

\vspace{5mm}

\noindent
Figure 9.\ \ $\chi^2$ as a function of $\Delta m^2$ along the $\sin^2
2\theta = 1$ line for the $\nu_{\mu} \to \nu_{\tau}$ scenario. Note the
shallow minimum. Note also that the minimum becomes
shallower still if $R$ is excluded from the fit. The insensitivity to
$\Delta m^2$ corresponds to the plateau features in Figs.2-6 (see text).

\vspace{5mm}

\noindent
Figure 10.\ \ The allowed region in the $(\sin^2 2\theta,\ \Delta m^2)$
plane for the $\nu_{\mu} \to \nu_{s}$ scenario.

\vspace{5mm}

\noindent
Figure 11.\ \ As for Figure 10 but with $R$ data excluded from the fit.

\vspace{5mm}

\noindent
Figure 12.\ \ As for Fig.9 but for the $\nu_{\mu} \to \nu_s$ scenario.

\newpage
\pagestyle{empty}
\unitlength=.2mm
\begin{center}
\begin{picture}(100,100)(0,450)
\put(-50,400){\makebox(100,100){\Huge{$\nu_\alpha$}}}
\put(70, 270){\makebox(100,100){\Huge{$\psi$}}}
\put(220, 220){\makebox(100,100){\Huge{$\alpha$}}}
\put(-90, 270){\makebox(100,100){\Huge{$\xi$}}}
\put(-150,230){\makebox(100,100){\Huge{$e_z$}}}
\put(10, 130){\makebox(100,100){\Huge{$\Theta$}}}
\put(-30, -130){\makebox(100,100){\Huge{$\phi$}}}
\put(-30, -430){\makebox(100,100){\Huge{\bf Fig.1}}}
\put(0,0){\vector(0,1){400}}
\put(0,0){\line(1,0){350}}
\put(0,0){\vector(1,1){235}}
\put(0,0){\vector(-1,3){81}}
\put(0,0){\line(2,-1){185}}
\put(185,185){\line(0,-1){277}}
\put(0,0){\line(-3,-2){200}}
\bezier{150}(0,300)(100,300)(185,185)
\bezier{150}(0,300)(-50, 260)(-60.,185)
\bezier{150}(185,185)(20,255)(-60.,185)
\bezier{150}(-39,-25)(0,-45)(50.,-25)
\end{picture}
\end{center}

\newpage
\epsfig{file=fig2.eps,width=15cm}
\newpage
\epsfig{file=fig3.eps,width=15cm}
\newpage
\epsfig{file=fig4.eps,width=15cm}
\newpage
\epsfig{file=fig5.eps,width=15cm}
\newpage
\epsfig{file=fig6.eps,width=15cm}
\newpage
\epsfig{file=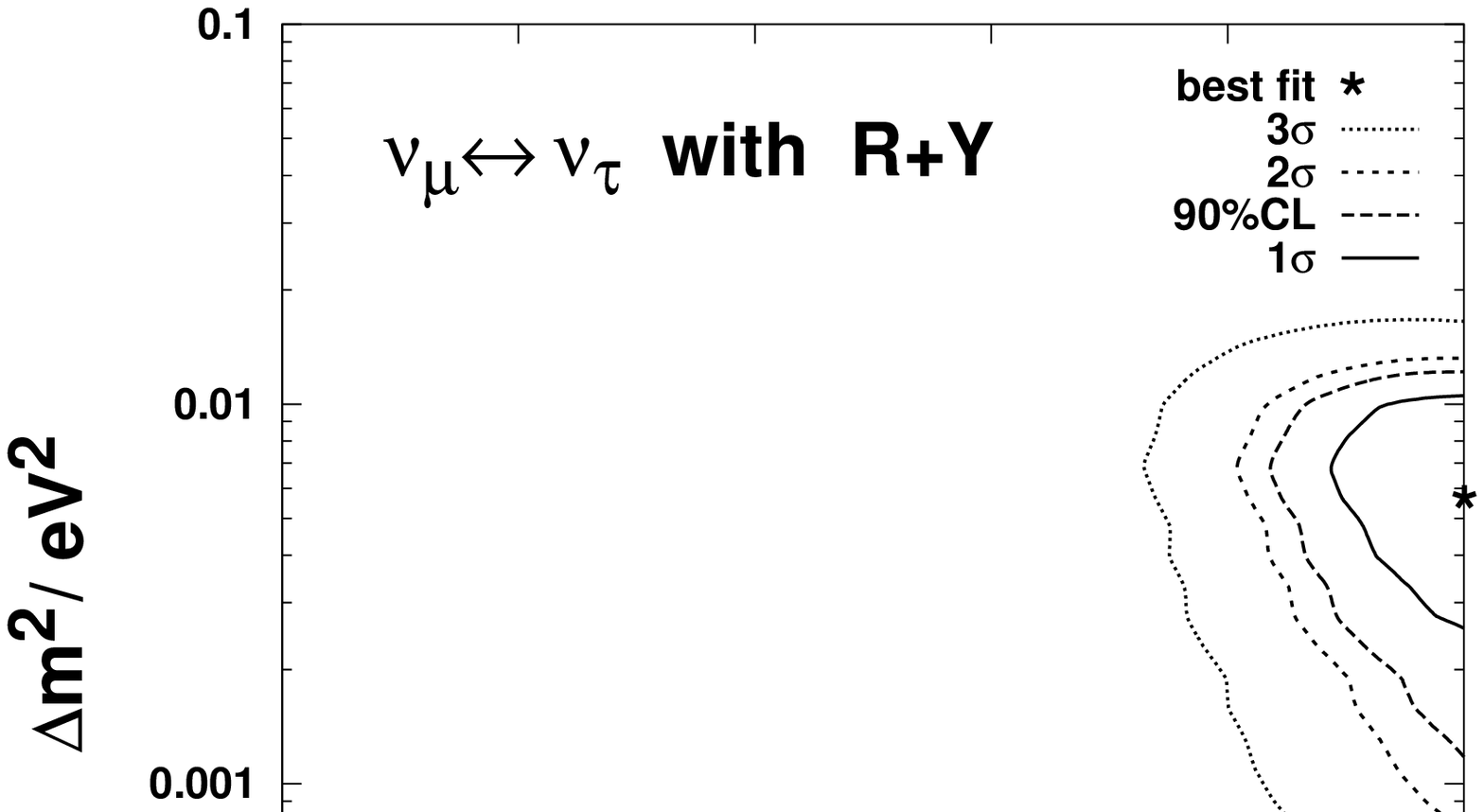,width=15cm}
\newpage
\epsfig{file=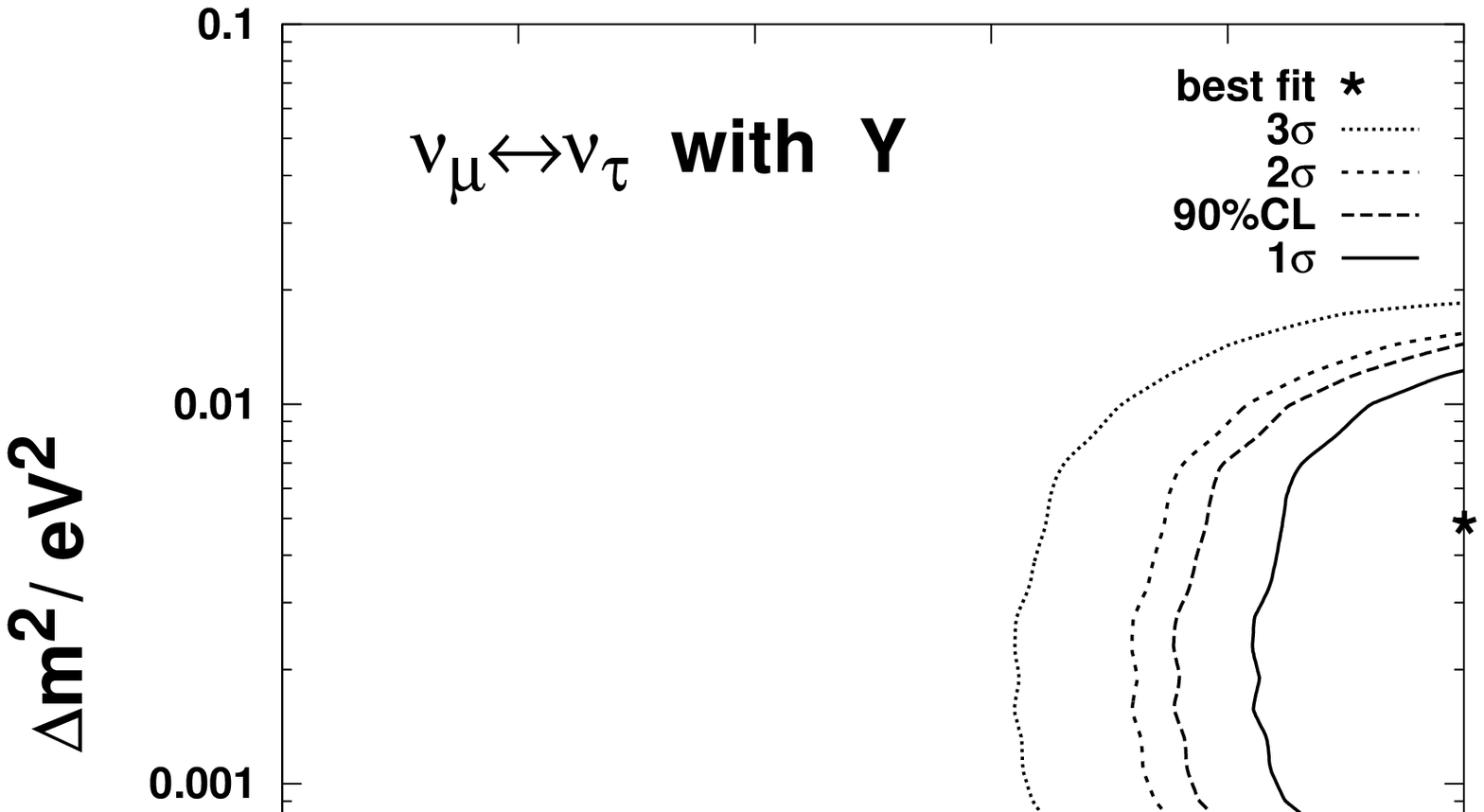,width=15cm}
\newpage
\epsfig{file=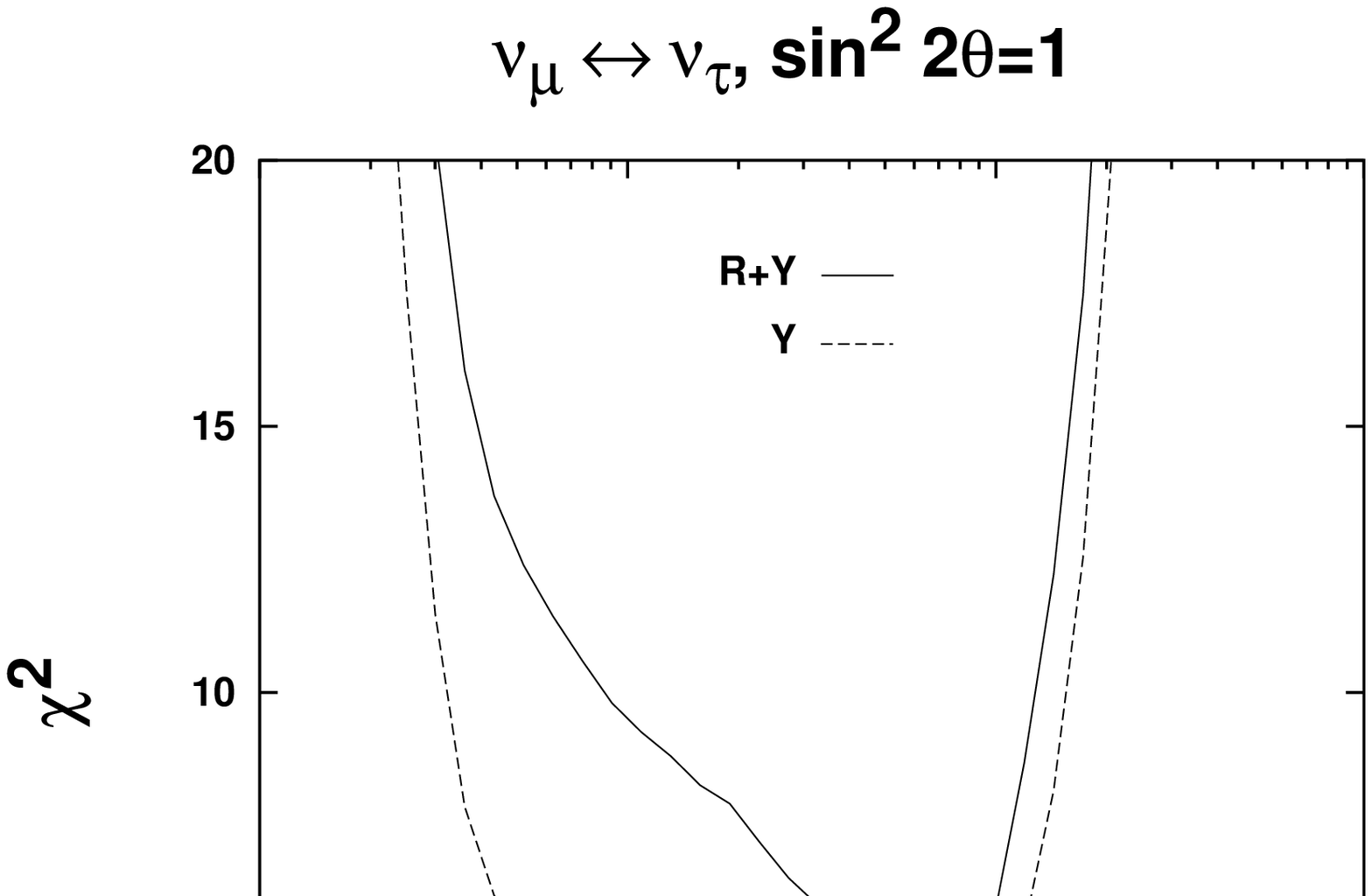,width=15cm}
\newpage
\epsfig{file=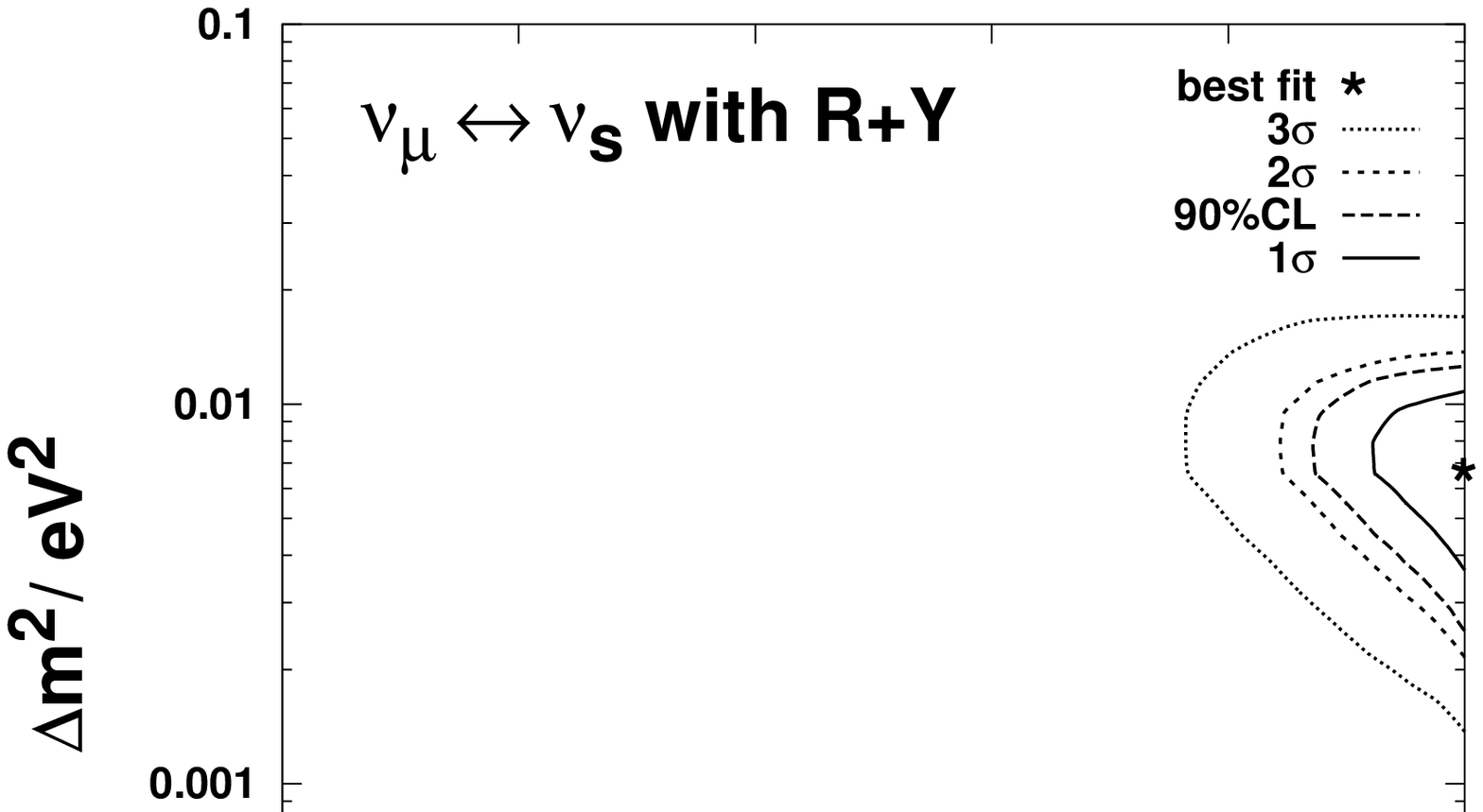,width=15cm}
\newpage
\epsfig{file=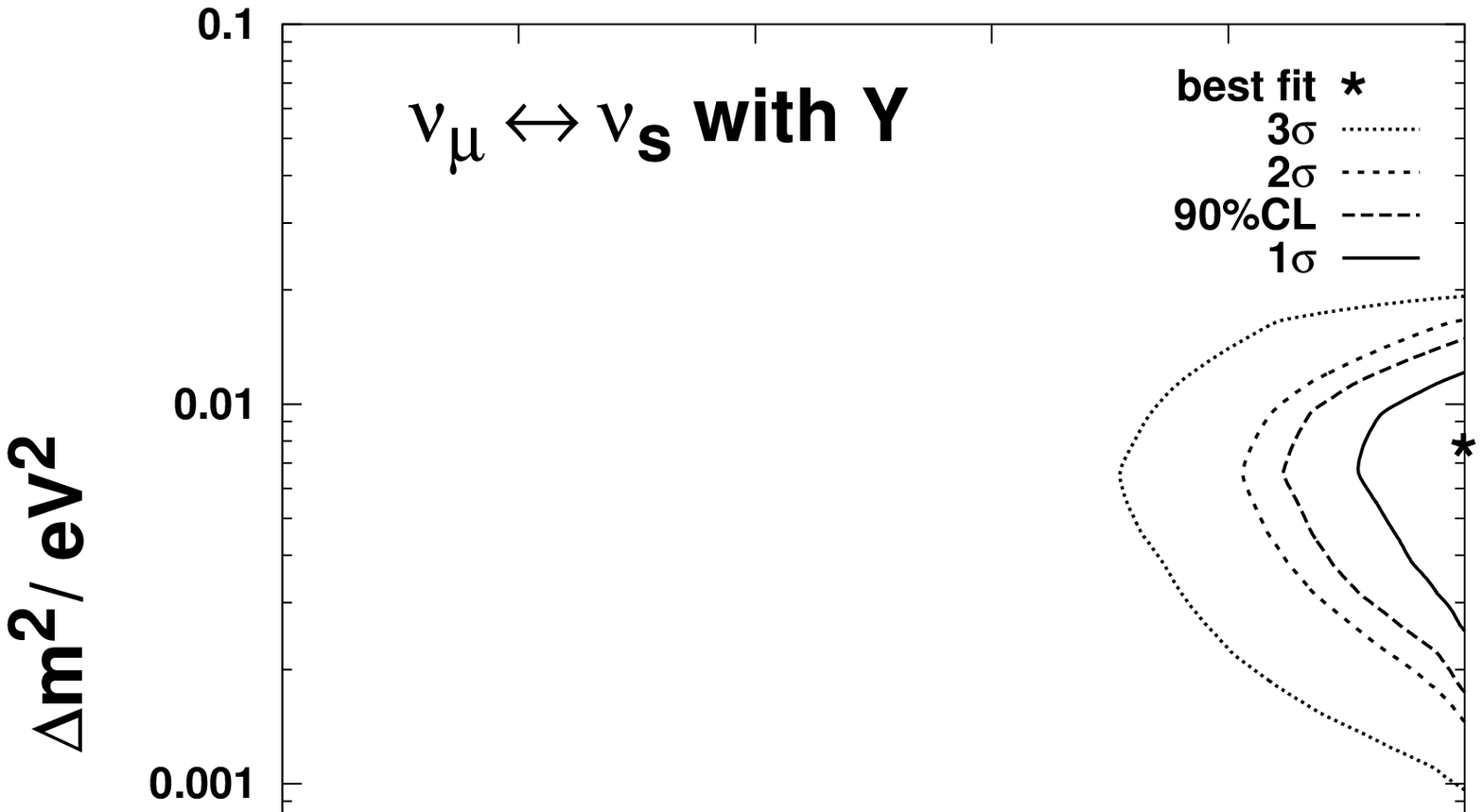,width=15cm}
\newpage
\epsfig{file=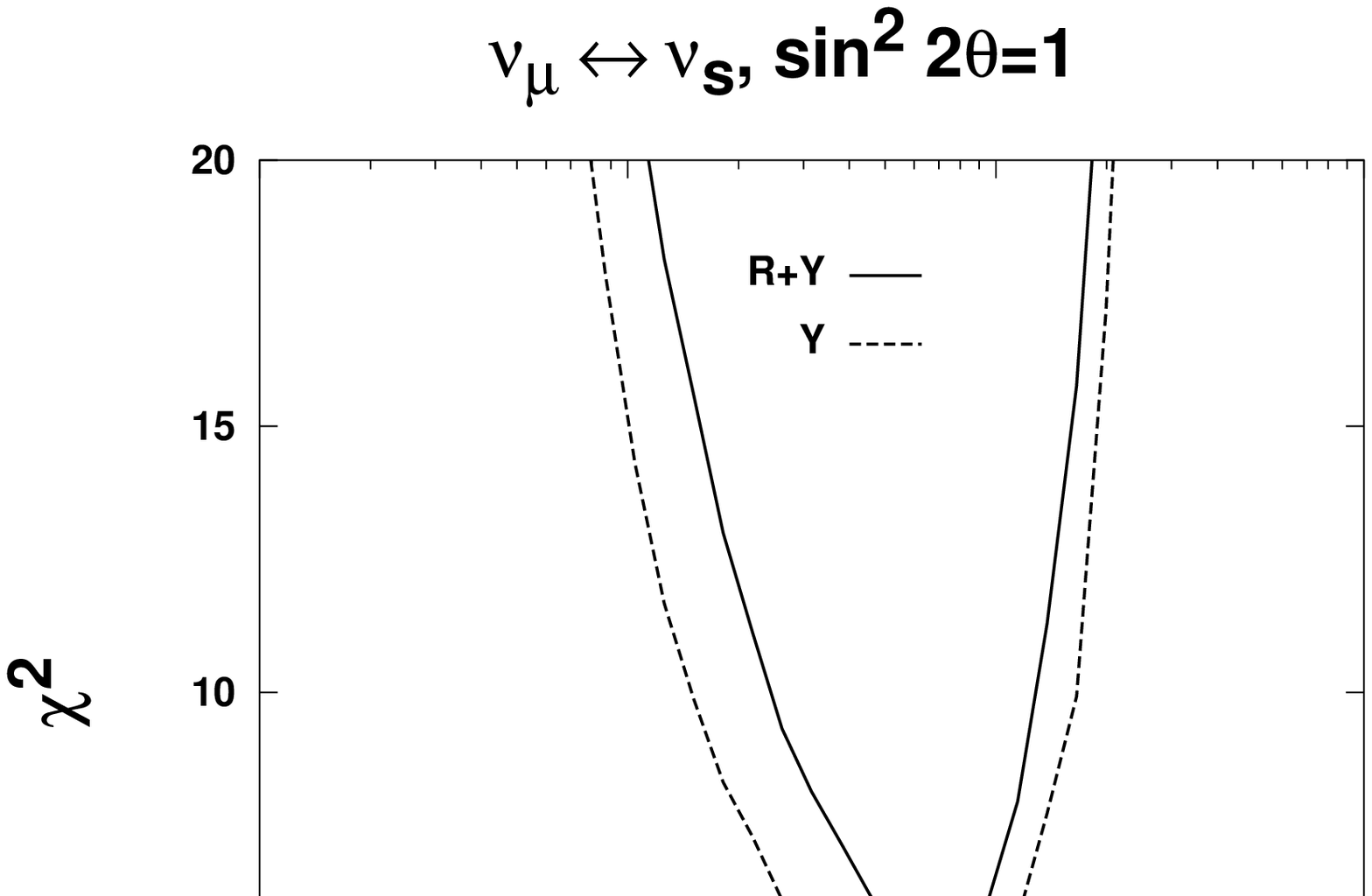,width=15cm}


\begin{thebibliography}{99} 

\bibitem{atmos}
Kamiokande Collaboration, K.S. Hirata et al.,  
Phys. Lett. {\bf B205}, 416 (1988);
{\it ibid.} {\bf B280}, 146 (1992);
Kamiokande Collaboration, Y. Fukuda et al., 
Phys. Lett. {\bf B335}, 237 (1994); 
IMB Collaboration, D. Casper et al., 
Phys. Rev. Lett. {\bf 66}, 2561 (1989);
R. Becker-Szendy et al., Phys. Rev. {\bf D46}, 3720 (1989);
NUSEX Collaboration, M. Aglietta et al., 
Europhys. Lett. {\bf 8}, 611 (1989);
Frejus Collaboration, Ch. Berger et al.,  
Phys. Lett. {\bf B227}, 489 (1989);
{\it ibid.} {\bf B245}, 305 (1990); 
K. Daum et al, Z. Phys. {\bf C66} 417 (1995); 
Soudan 2 Collaboration, M. Goodman et al., Nucl. Phys. {\bf B}
(Proc. Suppl.) {\bf 38},337 (1995);
W.W.M., Allison et. al., Phys. Lett. {\bf B391}, 491 (1997).

\bibitem{sk}
E. Kearns, Talk at {\it News about SNUS}, ITP Workshop, Santa Barbara,
Dec.\ 1997, http://doug-pc.itp.ucsb.edu/online/snu/kearns/oh/all.html.

\bibitem{mue}
A. Acker and S. Pakvasa, Phys. Lett. {\bf B397}, 209 (1997);
A. Acker, S. Pakvasa, J. Learned and T. J. Weiler, 
Phys. Lett. {\bf B298}, 149 (1993).

\bibitem{mutau}
J. G. Learned, S. Pakvasa and T. J. Weiler, 
Phys. Lett. {\bf B207},
79 (1988); V. Barger and K. Whisnant, Phys. Lett. 
{\bf B209}, 365 (1988);
K. Hidaka, M. Honda and S. Midorikawa, 
Phys. Rev. Lett. {\bf 61}, 1537 (1988);
J. T. Peltoniemi and J. W. F. Valle, Nucl. Phys. {\bf B406}, 409 (1993);
D. O. Caldwell and R. N. Mohapatra, Phys. Rev. {\bf D50}, 3477 (1994);
C. Y. Cardall and G. Fuller, Phys. Rev. {\bf D53}, 4421 (1996).

\bibitem{mus}
E. Akhmedov, P. Lipari and M. Lusignoli, Phys. Lett. {\bf B300}, 128 
(1993);
R. Foot, Mod. Phys. Lett. {\bf A9}, 169 (1994); 
R. Foot and R. R. Volkas, Phys. Rev. {\bf D52}, 6595 (1995);
Q. Y. Liu and A. Yu. Smirnov, hep-ph/9712493.
Note that the implications of the $\nu_{\mu} - \nu_s$ solution
for Big Bang Nucleosynthesis for the case of maximal mixing are studied
in R. Foot and R. R. Volkas,
Phys. Rev. {\bf D55}, 5147 (1997); Astropart. Phys. {\bf 7}, 283 (1997);
Phys. Rev. {\bf D56}, 6653 (1997).
For other
models with large angle or maximal active-sterile mixing see, for instance,
M. Kobayashi, C. S. Lim and M. M. Nojiri,
Phys. Rev. Lett. {\bf 67}, 1685 (1991); 
C. Giunti, C. W. Kim and U. W. Lee,
Phys. Rev. {\bf D46}, 3034 (1992); J. Bowes and R. R. Volkas,
University of Melbourne Preprint UM-P-97/09.

\bibitem{3fl}
In this paper we focus on 2 flavour oscillations. For 3 flavour analyses of
the atmospheric neutrino problem see, for example,  O. Yasuda,
TMUP-HEL-9706, hep-ph/9706546; S. M.
Bilenky, C. Giunti and C. W. Kim, Astropart. Phys. {\bf 4}, 241 (1996); G.
L. Fogli, E. Lisi, D. Montanino and G. Scioscia, Phys. Rev. {\bf D55}, 4385
(1997); M. Narayan, G. Rajasekaran and S. Uma Sankar, Phys. Rev. {\bf D56},
437 (1997); P. H. Harrison, D. Perkins and W. G. Scott, Phys. Lett. B396,
186 (1997); {\bf B349}, 137 (1995).

\bibitem{chooz}
CHOOZ Collaboration, M. Apollonio et al., hep-ex/9711002.

\bibitem{msw}
L. Wolfenstein, Phys. Rev. {\bf D17}, 2369 (1978); S. P. Mikheyev and A. Yu
Smirnov, Yad. Fiz. {\bf 42}, 1441 (1985) [Sov. J. Nucl. Phys. {\bf 42}, 913
(1985)]; Nuovo Cim. {\bf C9}, 17 (1986).

\bibitem{vissani}
T. Kajita, Talk at {\it Topical Workshop on Neutrino Physics},
Institute for Theoretical Physics, The University of Adelaide, Nov.\ 1996;
F. Vissani and A. Yu Smirnov, hep-ph/9710565.

\bibitem{liu}
See the paper by Liu and Smirnov in Ref.\cite{mus}.

\bibitem{macro}
F. Ronga, {\it Proceedings of the 17th International Conference on Neutrino
Physics and Astrophysics} (World Scientific, Singapore, 1997), eds. K.
Enqvist et al., p.529.

\bibitem{bla}
See for example,
D. Notzold and G. Raffelt, Nucl. Phys. {\bf B307}, 924 (1988).

\bibitem{earth}
F. Stacey, {\it Physics of the Earth, 2nd ed.} (J. Wiley and Sons,
Chichester, 1977). 

\bibitem{gaisser}
T. K. Gaisser and J. S. O'Connell, Phys. Rev. {\bf D34}, 822 (1986).

\bibitem{barger}
V. Barger and R. J. N. Phillips, {\it Collider Physics} (Addison-Wesley,
Redwood City, 1987), p.153.

\bibitem{hkkm}
M. Honda, T. Kajita, S. Midorikawa, and K. Kasahara,
Phys. Rev. {\bf D52}, 4985 (1995).

\bibitem{hkm}
M. Honda, K. Kasahara, and S. Midorikawa, private communication.

\bibitem{flux}
L.V. Volkova, Sov. J. Nucl. Phys. {\bf 31}, 784 (1980);
T.K. Gaisser, T. Stanev S.A. Bludman and H. Lee, Phys. Rev. Lett.
{\bf 51}, 223 (1983);
A. Dar, Phys. Rev. Lett.
{\bf 51}, 227 (1983);
K. Mitsui, Y. Minorikawa and H. Komori, Nuovo Cim. {\bf C9}, 995 (1986);
E.V. Bugaev and V.A. Naumov, Sov. J. Nucl. Phys. {\bf 45}, 857 (1987);
T.K. Gaisser, T. Stanev and G. Bar, Phys. Rev. {\bf D38}, 85 (1988);
A.V. Butkevich, L.G. Dedenko and I.M. Zheleznykh, Sov. J. Nucl. Phys.
{\bf 50}, 90 (1989);
M. Honda, K. Kasahara, K. Hidaka and S. Midorikawa, Phys. Lett. {\bf B248},
193 (1990);
H. Lee and Y. S. Koh, Nuovo Cim. {\bf B105}, 883 (1990);
M. Kawasaki and S. Mizuta, Phys. Rev. {\bf D43}, 2900 (1991);
P. Lipari, Astropart. Phys. {\bf 1}, 195 (1993).
D.H. Perkins, Astropart. Phys. {\bf 2}, 249 (1994);
V. Agrawal, T.K. Gaisser, P. Lipari and T. Stanev,
Phys. Rev. {\bf D53}, 1314 (1996); T.K. Gaisser and T. Stanev,
astro-ph/9708146.

\bibitem{updown}
J. Bunn, R. Foot and R. R. Volkas, Phys. Lett. {\bf B413}, 109 (1997);
J. W. Flanagan, J. G. Learned and S. Pakvasa, hep-ph/9709438;
R. Foot, R. R. Volkas and O. Yasuda, hep-ph/9709483, Phys. Rev. {\bf D} (in
press); hep-ph/9710403, Phys. Lett. {\bf B} (in press).

\bibitem{fn}
In our numerical work we found some small oscillations ($\sim 5\%$) of
the $Y$ parameter which we have smoothed out in the Figures.

\bibitem{fogli}
G. L. Fogli and E. Lisi, Phys. Rev. {\bf D52}, 2775 (1995).

\bibitem{fit}
M. C. Gonzalez-Garcia et al., hep-ph/9801368.

\end{thebibliography}
\end{document}